\documentclass[conference]{IEEEtran}
\IEEEoverridecommandlockouts
\usepackage{cite}
\usepackage{amsmath,amssymb,amsfonts}
\usepackage{algorithmic}
\usepackage{graphicx}
\usepackage{textcomp}
\usepackage{tcolorbox}
\usepackage{verbatim}
\usepackage{soul,color}
\usepackage[table,xcdraw]{xcolor}
\usepackage{multirow}
\usepackage[linesnumbered,ruled,vlined]{algorithm2e}

\def\BibTeX{{\rm B\kern-.05em{\sc i\kern-.025em b}\kern-.08em
    T\kern-.1667em\lower.7ex\hbox{E}\kern-.125emX}}
\begin{document}

\title{Tailored Immersive Environments: Advancing Neurodivergent Support Through Virtual Reality\\
}

\author{
\IEEEauthorblockN{Elia Moscoso-Thompson}
\IEEEauthorblockA{CNR-IMATI\\Genoa, Italy\\
0000-0003-1230-8291}
\and
\IEEEauthorblockN{Katia Lupinetti}
\IEEEauthorblockA{CNR-IMATI\\
Genoa, Italy\\0000-0002-0202-4909}
\and
\IEEEauthorblockN{Irene Capasso}
\IEEEauthorblockA{UNIGE-DIBRIS\\Genoa, Italy\\
0009-0005-1562-999X}
\and
\IEEEauthorblockN{Fabrizio Ravicchio}
\IEEEauthorblockA{CNR-ITD\\Genoa, Italy\\
0000-0002-1344-2209}
\and
\IEEEauthorblockN{Brigida Bonino}
\IEEEauthorblockA{CNR-IMATI\\Genoa, Italy\\
0000-0002-4264-3958}
\and
\IEEEauthorblockN{Franca Giannini}
\IEEEauthorblockA{CNR-IMATI\\Genoa, Italy\\
0000-0002-3608-6737}
\and
\IEEEauthorblockN{Andrea Canessa}
\IEEEauthorblockA{UNIGE-DIBRIS\\Genoa, Italy\\
0000-0001-8946-5290}
\and
\IEEEauthorblockN{Silvio Sabatini}
\IEEEauthorblockA{UNIGE-DIBRIS\\Genoa, Italy\\
0000-0002-0557-7306}
\and
\IEEEauthorblockN{Lucia Ferlino}
\IEEEauthorblockA{CNR-ITD\\Genoa, Italy\\
0000-0002-8021-1793}
\and
\IEEEauthorblockN{Chiara Malagoli}
\IEEEauthorblockA{CNR-ITD\\Genoa, Italy\\
0000-0001-6872-2315}
}

\maketitle

\begin{abstract}
Every day life tasks can present significant challenges for neurodivergent individuals, particularly those with Autism Spectrum Disorders (ASD) who are characterized by specific sensitivities. This contribution describes a virtual reality system that allows neurodivergent individuals to experience everyday situations in order to practice and implement strategies for overcoming their daily challenges. The key strength of the proposed system is the automatic personalization of the virtual environment, based on both the individual's abilities and their specific training needs. The proposed method has been evaluated on four synthetic user profiles, also proposing a metric able to evaluate the variance of the features within the same difficulty level.
The results show that the method can produce a significant number of scenarios for the various difficulty levels. Furthermore, within the same difficulty, there is a wide variance of the non-constrained features for the specific profile.
\end{abstract}


\section{Introduction}
Urbanisation is advancing at an unprecedented rate, expanding city boundaries and population density. While new technologies make modern cities more connected, the resulting sensory and social complexity can hinder mobility, especially for people whose perceptual and cognitive processing differs from the neurotypical norm. Among them, neurodivergent individuals are disproportionately affected: atypical sensory filtering and heightened cognitive load interact with dense traffic, unpredictable pedestrian behaviour, and highly stimulating architecture, turning apparently mundane journeys into demanding problem-solving tasks.

Current therapeutic approaches for individuals with neurodivergent conditions are often unable to integrate the sensory and perceptual complexity of real-world environments. This limits the real-life applicability and generalization of coping strategies learned during therapy.
From the research point of view, new technologies are employed to improve current approaches and 
 virtual reality applications has been investigated deeply \cite{abich2021review}.
Didehbani et al. \cite{ke2022virtual} proposed a VR training through a desktop application that allows trainees to test their abilities on 10 prefabricated modules, interacting with virtual avatars. 
Frolli et al. \cite{frolli2022children} proposed an emotional training in virtual reality and compared it with traditional emotional training performed individually with a therapist. 
Saiano et al. \cite{saiano2015natural} proved that a gesture-based “natural” interface combined with a desktop virtual environment can safely train adults on the autism spectrum to follow paths and cross streets. 
These systems, although innovative from the technological point of view, are limited considering their personalization capabilities; indeed few adjustments are possible (not even in all the systems), and the tuning of the configurations strongly depends on the therapist. This limitation strongly prevents the possibility of tailoring the training to each individual’s specific needs.

To address this gap, we design and implement EASE VR—Empowering Accessible Social Engagement in Virtual Reality, a system for creating ecologically valid, personalized virtual environments that reflect the user's sensory and cognitive challenges.
Through EASE VR, users can face complex crossing situations in a controlled and safe environment. Each scenario can be adjusted to the sensory and cognitive needs of the individual through a therapist-facing personalization interface.


\section{The EASE VR system}\label{sez:vrsystem}
The EASE VR system includes distinct training scenarios conceptualized and designed to simulate real-world situations that can be perceived as particularly challenging for neurodivergent individuals. Each scenario allows participants to complete specific tasks that assess their ability in living different places of the city, and among the scenarios that EASE VR proposes, the \textbf{Urban Crosswalk Scenario} is particularly relevant, and in this work, we focus on this scenario and its personalization. It is designed to reproduce, through hyperrealistic graphics and high visual realism~\cite{Slater2009}, everyday life challenges for individuals with ASD connected to the extreme complexity of the city environment linked to sensory overstimulation~\cite{Miller2020} and the unpredictable dynamic of multiple situations and conditions (e.g., lights and weather conditions, noises, the number of people and vehicles, people's interaction, different waiting times, detours, unexpected behavior of others, vehicles passing despite stop signs, strikes, etc.). Consequently, this scenario integrates specific challenges such as managing waiting times, visuo-spatially orienting in the virtual environment while correctly assessing the situation, and extracting or requesting information to correctly manage the specific presented challenge, while dealing with sensory and social stimulation, providing integrated ecologically comparable physical and social functional mediators to support the ecological generalization of developed behaviors in real life~\cite{Stokes2016}.

In this scenario, participants must successfully cross a virtual street while adhering to standard pedestrian rules to reach a specific place. Depending on the difficulty level, the pedestrian crossing length and difficulty environmental conditions, such as time of day (day/night), weather (rain/sun), traffic density, and pedestrian flow can be modified and personalized according to the user's characteristics. Unexpected events, such as a vehicle running a red light or an ambulance passing through the intersection with sirens on, are introduced to experience and evaluate the participant’s ability to react to unforeseen situations, observing the ability of participants to manage the cognitive load and overall sensory stimulation.
An increasing complexity results in progressive modifications to the above-mentioned environmental parameters within the scenario, accompanied by alterations in stimulus presentation that respond dynamically to participant's performance and needs. This tailoring is individually calibrated both in terms of cognitive characteristics and level of functioning of each participant. The customization is facilitated through an external graphic user interface, which enables to adjust every element in the scenario. This means that the maximum difficulty threshold exhibits inter-individual variation, reflective each participant's sensory sensitivities and cognitive profile. 

\section{Personalization method and experiment setup}\label{sez:method}
According to the Urban Crosswalk Scenario design, the proposed personalization method enables the generation of VR environments, where the configurable features are:
\begin{itemize}
    \item \textit{Type of crossing}: short, long, double;
    \item \textit{Light setting}: day, night;
    \item \textit{Rain}: sunny, rain;
    \item \textit{Presence of pedestrians}: low, medium, high flow;
    \item \textit{Presence of vehicles}: low, medium, high flow;
    \item \textit{Sudden sound distractors}: presence or absence of church bells, helicopters, car red lights;
    \item \textit{Background noise}: volume noise of ambulances, baby crying, and dogs' barking;
    \item \textit{Type of traffic light}: absence, standard, presence of push-button and/or countdown, or broken.
\end{itemize}

\begin{figure*}[tbp]\centering
    \begin{tabular}{ccc}  
        \includegraphics[width=0.26\linewidth]{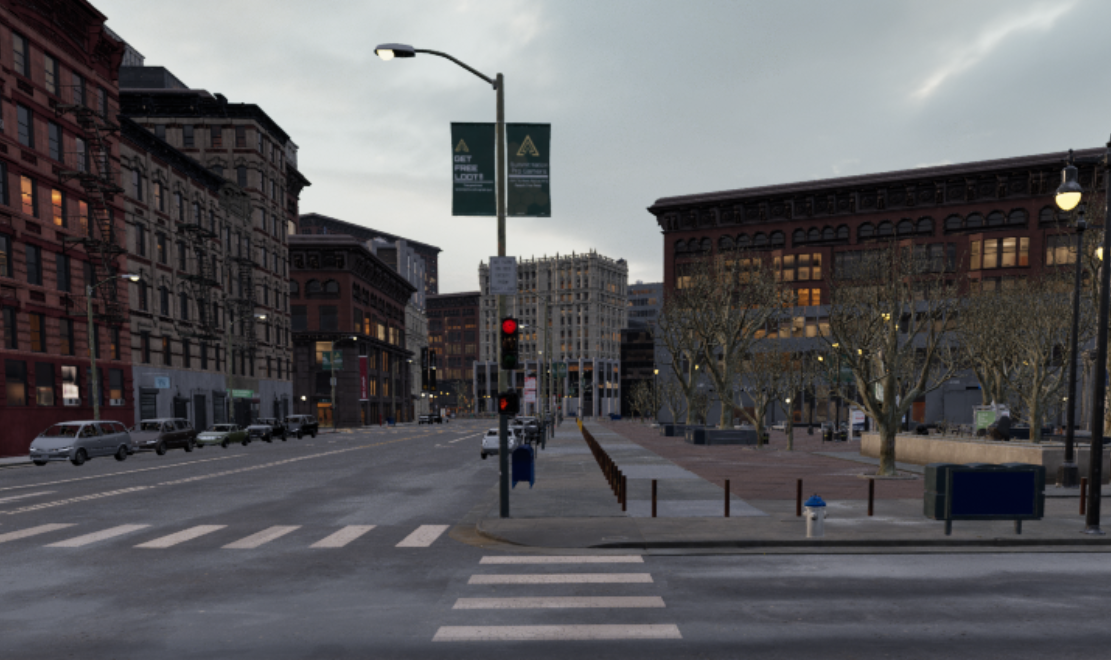} & 
        \includegraphics[width=0.26\linewidth]{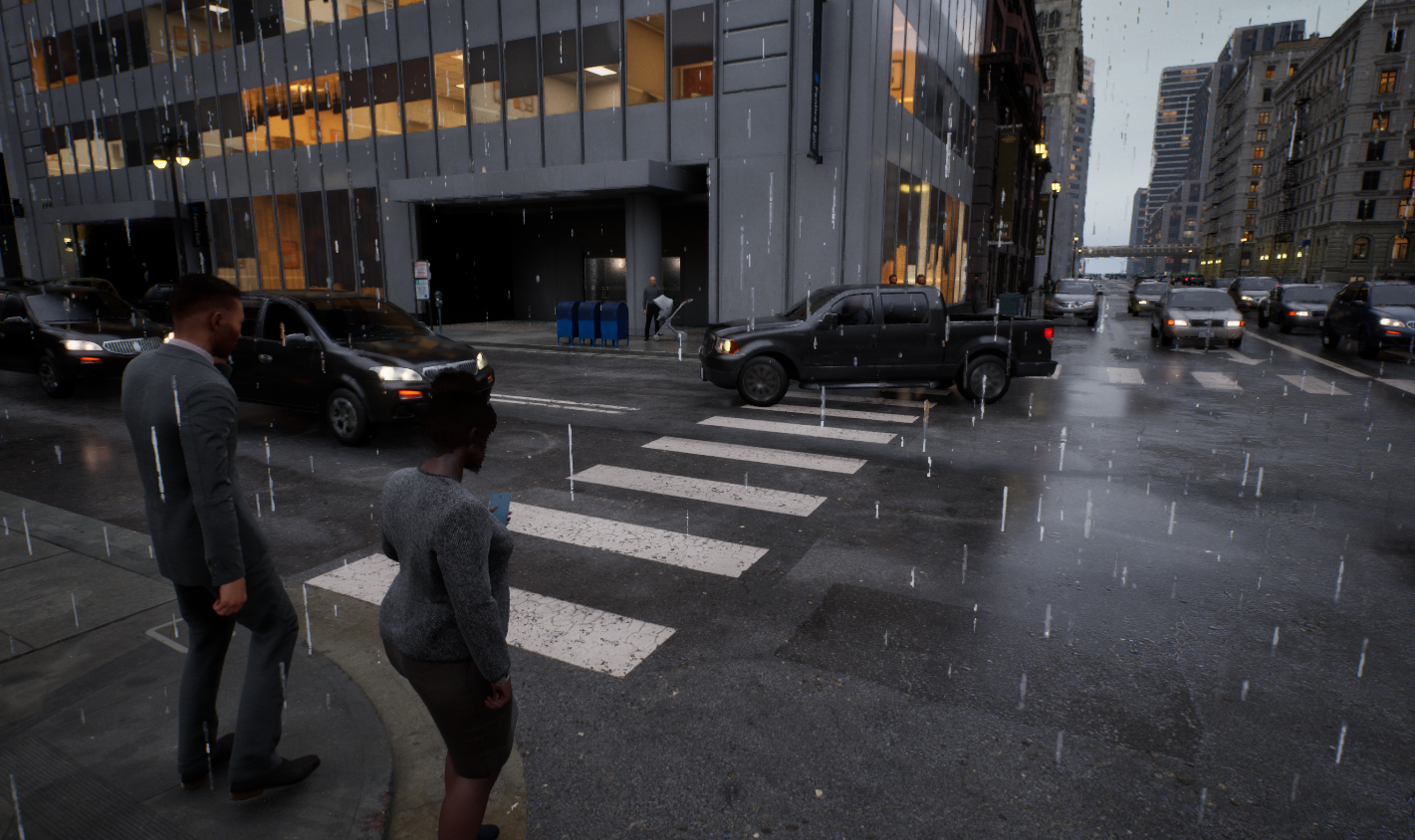}  &
        \includegraphics[width=0.26\linewidth]{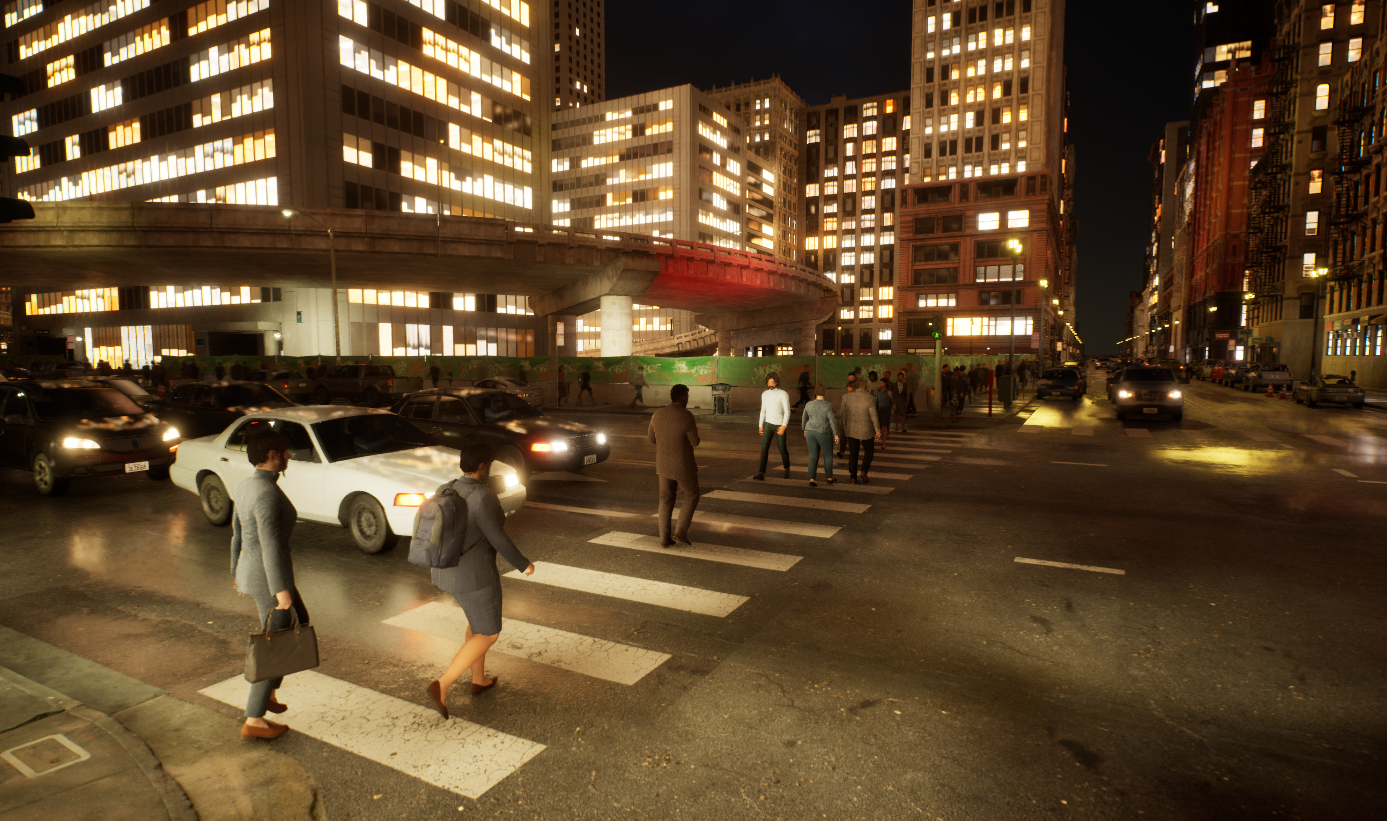} \\
        (a) Simple urban scenario & (b) Rainy urban scenario & (c) Night urban scenario
    \end{tabular}
    \caption{Examples of different training scenarios.} 
    \label{fig:ScenarioExamples}
\end{figure*}

Fig~\ref{fig:ScenarioExamples} presents three examples of pedestrian crossing scenarios obtained by combining various parameters previously described.
A value is assigned to each feature state, sorting them based on the expected level of difficulty. Without practical knowledge of the actual perceived difficulty, the values are equally spaced on the range $[0,1]$. For example, \textit{Type of crossing} can assume the value $1/3$ (short), $2/3$ (long), $1$ (double). 

To help formalize our analysis, we use $\phi_i$ to represent a feature. Moreover, to encode the different difficulty perceptions based on the user capabilities, we add a weight $\omega_j$ to each feature, ranging from $1$ to $5$. We group some features based on the \emph{aspect} we impact in the scenario, and we assign them the same weights (e.g., all \textit{Sudden sound distractors} have the same weight). For an overview of the notation used in this work, we refer to Table~\ref{tab:notation}. 

\begin{table*}[ht]
\centering
\caption{Overview of the notation used in the analysis to describe a scenario.}
\label{tab:notation}
\begin{tabular}{|cl|c|cl|}
\hline
\multicolumn{2}{|c|}{Feature name} & \multicolumn{1}{c|}{Feature values} &                          \multicolumn{2}{|c|}{Cognitive skill}                          \\ \hline
\rowcolor[HTML]{EFEFEF} 
$\phi_1$                               & Type of crossing                  & \{1/3 (short), 2/3 (long), 1 (double)\}                     & $\omega_1$                                           & Visuospatial awareness \\
$\phi_2$                               & Night time                        & \{0 (day time), 1 (night time)\}                            &                                                      &                                                    \\
$\phi_3$                               & Rain                              & \{0 (sunny), 1 (rainy)\}                            & \multirow{-2}{*}{$\omega_2$}                         & \multirow{-2}{*}{Pattern vision}                        \\
\rowcolor[HTML]{EFEFEF} 
$\phi_4$                               & Presence of pedestrians           & \{0 (no one), 1/2 (some people), 1 (many people)\}                       & $\omega_3$                                           & Social factor                         \\
$\phi_5$                               & Presence of vehicles              & \{0 (no cars), 1/2 (some cars), 1 (many cars)\}                       & $\omega_4$                                           & Hazard factor                            \\
\rowcolor[HTML]{EFEFEF} 
$\phi_6$                               & ssd: church bell                  & \{0 (no sound), 1 (sound activated)\}                            & \cellcolor[HTML]{EFEFEF}                             & \cellcolor[HTML]{EFEFEF}                      \\
\rowcolor[HTML]{EFEFEF} 
$\phi_7$                               & ssd: helicopter                   & \{0 (no sound), 1 (sound activated)                            & \cellcolor[HTML]{EFEFEF}                             & \cellcolor[HTML]{EFEFEF}                        \\
\rowcolor[HTML]{EFEFEF} 
$\phi_8$                               & ssd: car wating red lights        & \{0 (no sound), 1 (sound activated)                            & \multirow{-3}{*}{\cellcolor[HTML]{EFEFEF}$\omega_5$} & \multirow{-3}{*}{\cellcolor[HTML]{EFEFEF}Sudden Sound perception} \\
$\phi_9$                               & bn: ambulance                     & \{0, 1/3, 2/3, 1\}*                 &                                                      &                                                                   \\
$\phi_{10}$                               & bn: baby crying                   & \{0, 1/3, 2/3, 1\}*                 &                                                      &                                                                   \\
$\phi_{11}$                               & bn: dogs barking                  & \{0, 1/3, 2/3, 1\}*                 & \multirow{-3}{*}{$\omega_6$}                         & \multirow{-3}{*}{Tolerance to noise}                                \\
\rowcolor[HTML]{EFEFEF} 
$\phi_{12}$                               & Traffic light                     & \{1/6, 1/3, 1/2, 2/3, 5/6, 1\}**    & $\omega_7$                                           & Rule complexity and hazard factor                         \\ \hline       
\multicolumn{5}{c}{}\\
\end{tabular}

\scriptsize{$(*)$: background noises (bn) varies from $0$ to $1$. Given the exploratory nature of this work, we only assume there are four possible volume setups: mute, low, medium, and high. $(**)$: we considered the following traffic light (TL) states:  (1) no traffic light, (2) presence of a basic TL, (3) TL with a pedestrian button, (4) TL with a visible countdown timer, (5) TL with both button and timer, and (6) broken or malfunctioning TL.}

\end{table*}

The personalization algorithm follows a four-step process to generate training scenarios aligned with user-specific profiles.
\begin{itemize}
    \item First, all possible combinations of scenario parameters are computed by systematically varying the values of each feature. This exhaustive enumeration ensures full coverage of the scenario space.
    \item Second, for each combination, a difficulty score is computed using a linear weighted sum:
    \begin{equation}
        D_{\text{score}} = \sum_{(i,j)} \omega_{i} \phi_{j},
    \end{equation}
    for opportune $(i,j)$ combinations based on what reported in Table~\ref{tab:notation}. This \emph{difficulty score} allows for ranking each scenario difficulty according to the specific profile (encoded through the weights). Notice that, if a feature can be disabled on a scenario (e.g., a sound), we also add $0$ to the possible values, so that the weight does not influence to the score if feature is not present.
    \item Third, profile-specific constraints are applied to filter out combinations that contain features deemed unsuitable or non-beneficial for the given user (e.g., suppressing loud stimuli for sound-sensitive users at the beginning of the training), or selecting exactly those elements that are relevant for the traing (e.g. a certain type of traffic light).
    \item Finally, the resulting difficulty scores are normalized to the $[0, 1]$ interval, enabling consistent comparisons across profiles and scenario sets.
\end{itemize}
This entire process exhibits linear computational complexity with respect to the number of combinations, making the method both scalable and computationally efficient for real-time scenario generation.
Here, the idea is that for a fixed difficulty score $D_{score}$, the generated scenarios vary in content while preserving the overall training challenge level. This enables the construction of training paths that are ecologically diverse but therapeutically consistent. In addition, to ensure clinical relevance, constraints are applied after scenario generation. These constraints enforce the presence (or exclusion) of specific features depending on the user profile.

\subsection{Score discretization and consistent difficulty}
Given the number of features and the range of values each feature can assume, the number of different scores varies significantly. To support the exploratory purpose of this work, we group similar scores with an adaptive discretization approach. In other words, we discretize the scenarios $D_{score}$ so that two different scores have significant differences in feature setup and value. 
Specifically, we define the larger possible score increase, denote it $\delta$, and use it to segment the score range $[0, 1]$. A more formal definition of $\delta$ is as follow:
\begin{equation}
    \delta = \text{max}(\omega_i\phi_i), \qquad \forall i,
\end{equation}
reminding that $\phi_i$ and $\omega_i$ represent the feature value and related weight, respectively.
This choice ensures that each score interval includes meaningful features variation.

Then, we cluster scenarios with similar $D_{\text{score}}$, referring to these cluster as \emph{scenarios sets with consistent difficulty} ($CD_{\text{score}}$). These are computed as follows:
\begin{equation}
    CD_{\text{score}} =round\left(\frac{D_{\text{score}}}{\delta}\right)^{-1},
\end{equation}
where $round(\cdot)$ denotes the round operator. Taking the reciprocal of the ratio helps normalize  $CD_{\text{score}}$ values to lie between $0$ and $1$.


\begin{figure*}[t]
    \centering
    \begin{tabular}{cc}
        \multicolumn{2}{c}{Profile 1: Sound hypersensitive (290304 out of 331776 profile specific scenarios - 87.5\%)}\\
        \includegraphics[height=3.8cm]{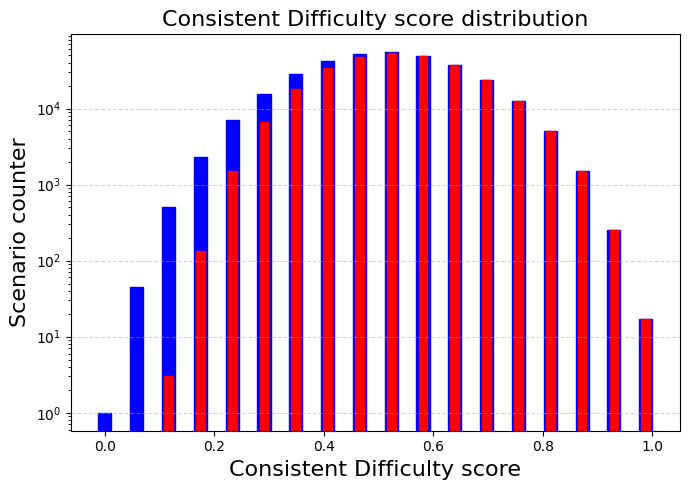} & 
        \includegraphics[height=3.8cm]{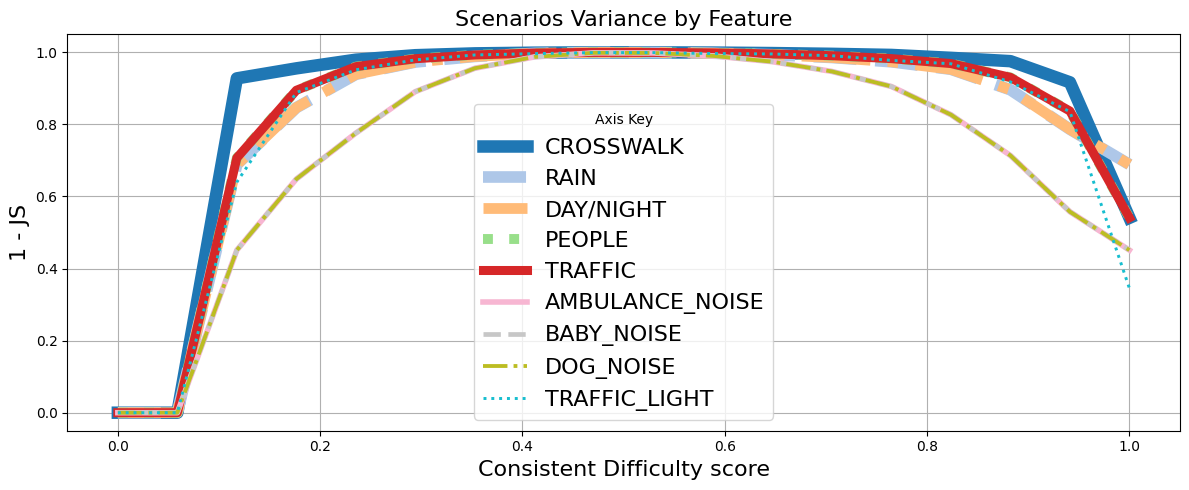}\\

        \begin{tabular}{
            >{\columncolor[HTML]{0000FF}}l ll
            >{\columncolor[HTML]{FE0000}}l l}
             & \scriptsize{All scenarios} &  & {\color[HTML]{FE0000} } & \scriptsize{Profile specific scenarios}
        \end{tabular}

    \end{tabular}
        
    \caption{Profile 3 scenarios analysis. Left: the distribution of scenarios (the y-axis is in logarithmic scale). Right: the related variance in features values.}
    \label{fig:pr1}
\end{figure*}

\begin{figure*}[t]
    \centering
    \begin{tabular}{cc}
        \multicolumn{2}{c}{Profile 2: Excessively focused attention on a detail }\\ \\
        \multicolumn{2}{c}{Vol $<33\%$ (9216 out of 331776 profile specific scenarios - 2.7\%)}\\
        \includegraphics[height=3.8cm]{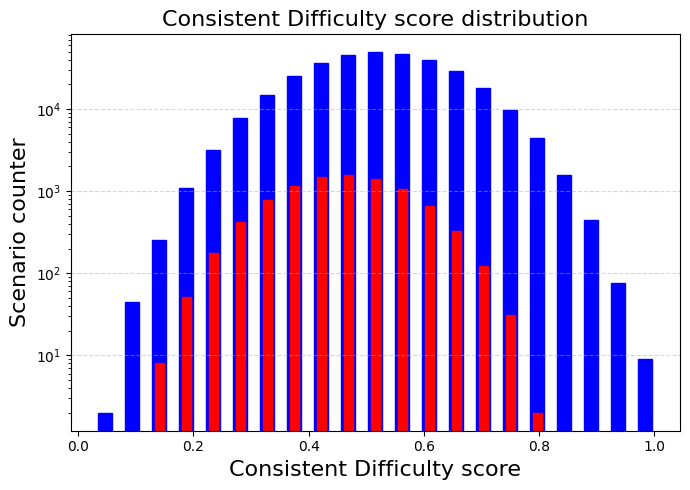} & 
        \includegraphics[height=3.8cm]{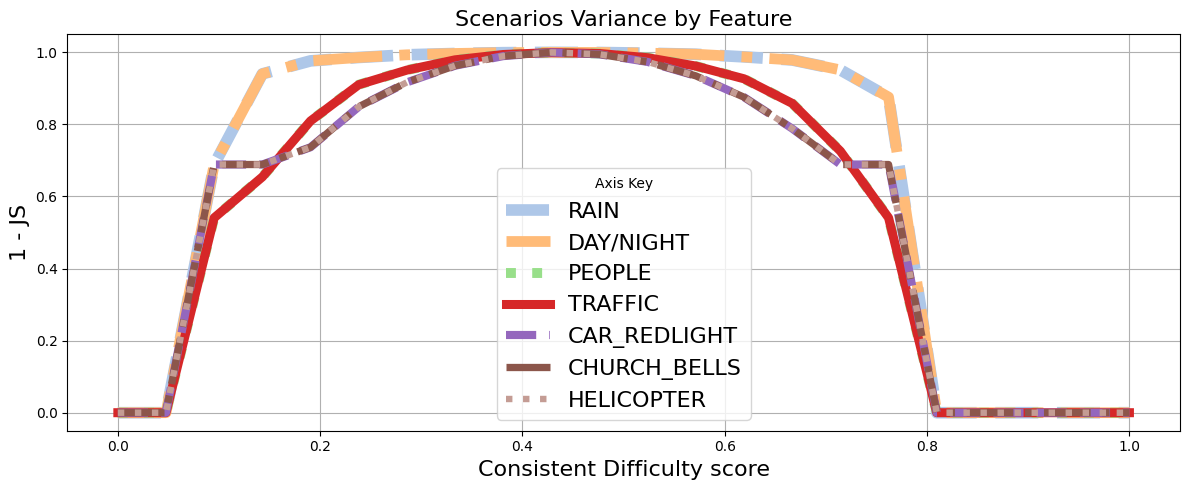}\\

        \multicolumn{2}{c}{$33\%<$ Vol $<66$ (31104 out of 331776 profile specific scenarios - 9.4\%)}\\
        \includegraphics[height=3.8cm]{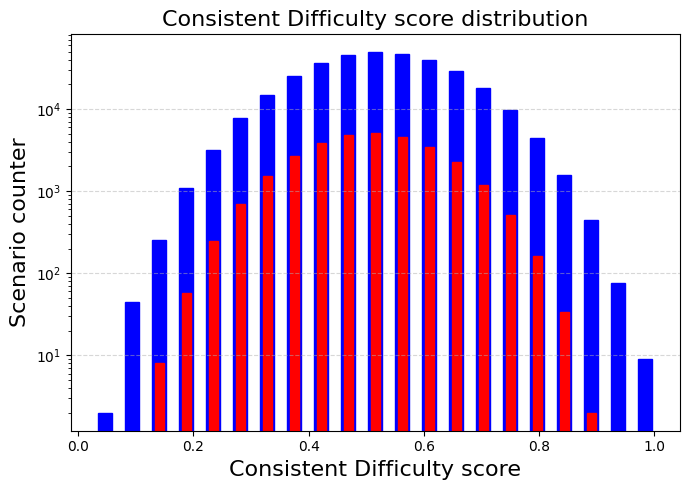} & 
        \includegraphics[height=3.8cm]{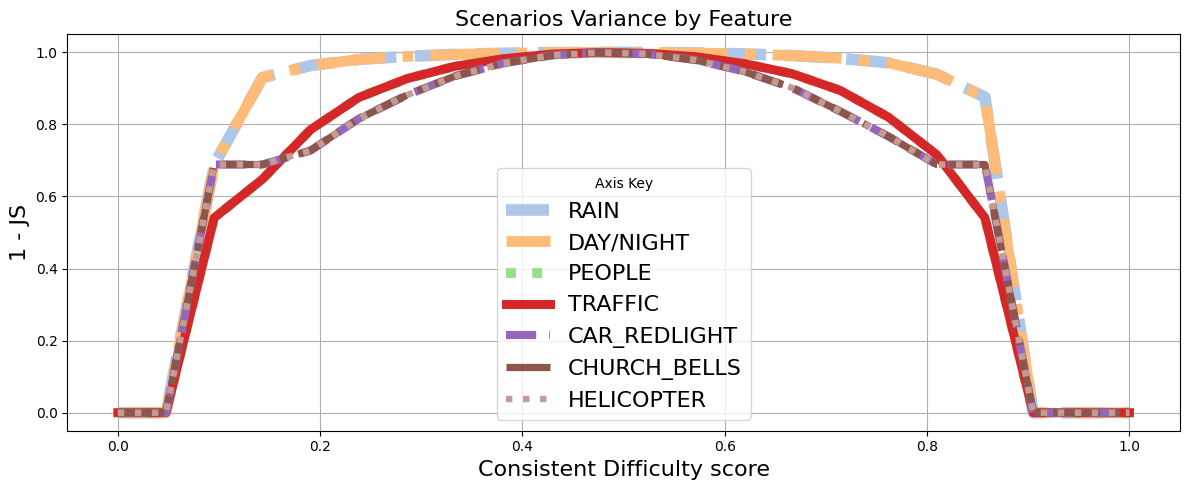}\\

        \multicolumn{2}{c}{$66\%<$ Vol $<100\%$ (73728 out of 331776 profile specific scenarios - 22.2\%)}\\
        \includegraphics[height=3.8cm]{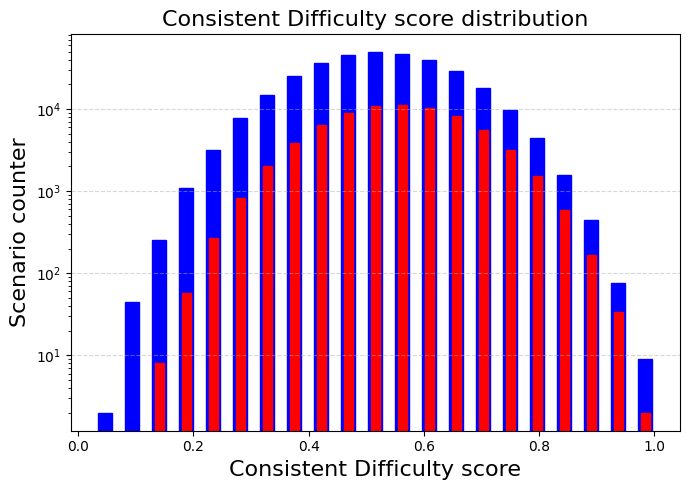} & 
        \includegraphics[height=3.8cm]{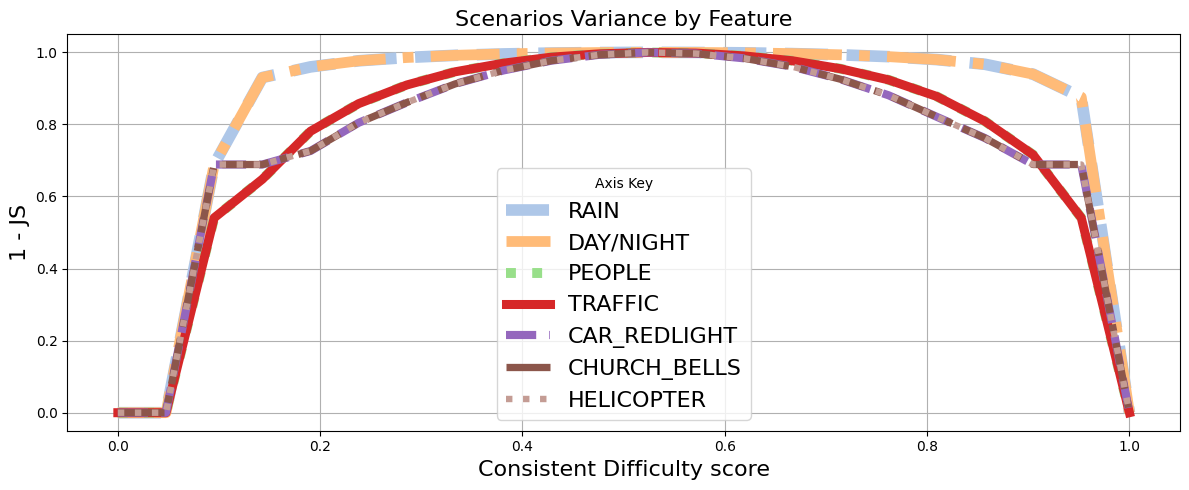}\\

        \begin{tabular}{
            >{\columncolor[HTML]{0000FF}}l ll
            >{\columncolor[HTML]{FE0000}}l l}
             & \scriptsize{All scenarios} &  & {\color[HTML]{FE0000} } & \scriptsize{Profile specific scenarios}
        \end{tabular}

    \end{tabular}
    
    \caption{Profile 2 scenarios analysis. Left: the distribution of scenarios (the y-axis is in logarithmic scale). Right: the related variance in features values.}
    \label{fig:pr2}
\end{figure*}

\begin{figure*}[ht]
    \centering
    \begin{tabular}{cc}
        \multicolumn{2}{c}{Profile 3: Social anxiety (16384 out of 331776 profile specific scenarios - 4.9\%)}\\
        \includegraphics[height=3.8cm]{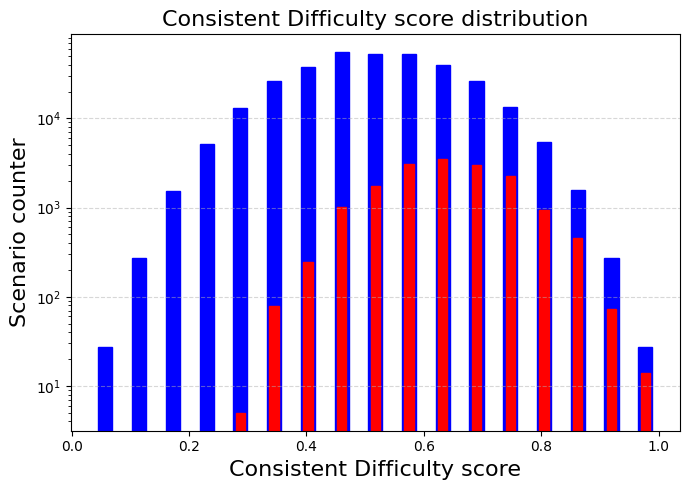} & 
        \includegraphics[height=3.8cm]{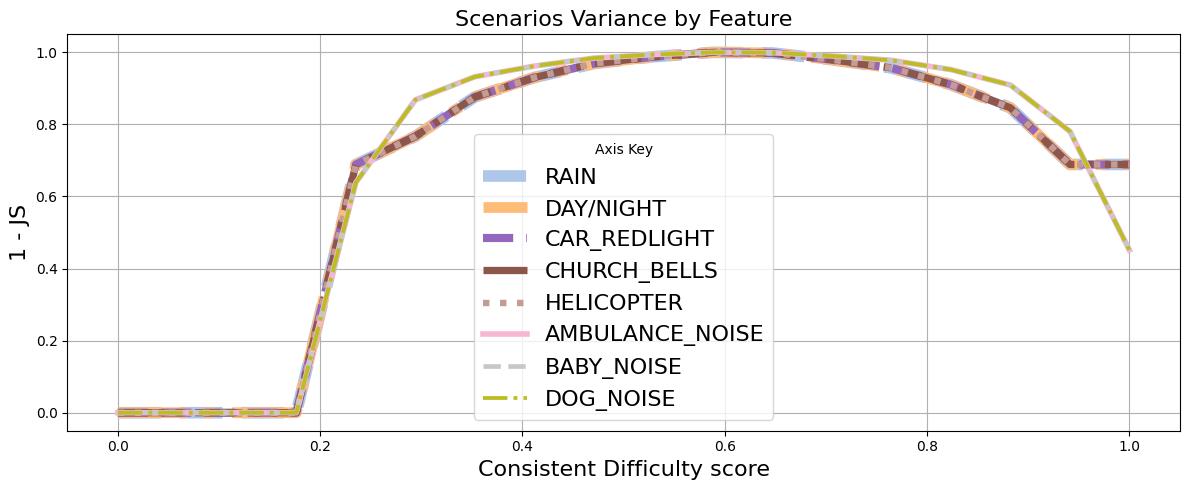}\\
        \\

         \multicolumn{2}{c}{Profile 4: Intermittent attention (147456 out of 331776 profile specific scenarios - 44.4\%)}\\
         \includegraphics[height=3.8cm]{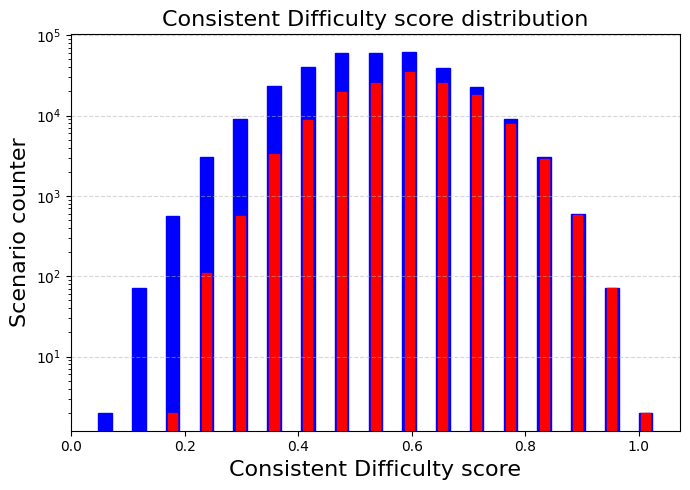} & 
         \includegraphics[height=3.8cm]{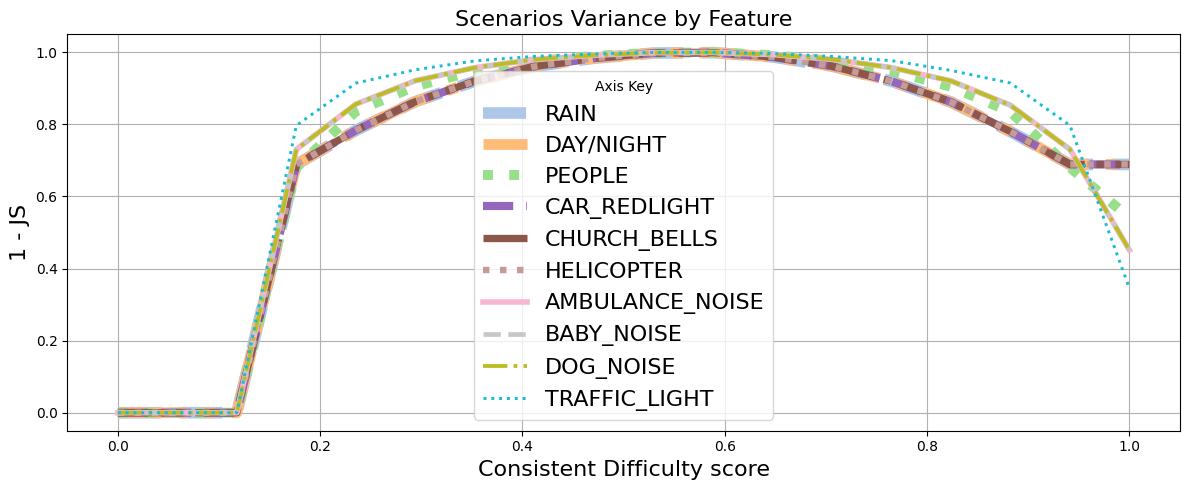}\\

         \begin{tabular}{
             >{\columncolor[HTML]{0000FF}}l ll
             >{\columncolor[HTML]{FE0000}}l l}
              & \scriptsize{All scenarios} &  & {\color[HTML]{FE0000} } & \scriptsize{Profile specific scenarios}
         \end{tabular}
     \end{tabular}
     \caption{Profile 3 and 4 scenarios analysis. Left: the distribution of scenarios (the y-axis is in logarithmic scale). Right: the related variance in features values.}
     \label{fig:pr4}
 \end{figure*}
\section{Evaluation objectives and adopted metrics}
\label{sez:evaluation_results}
This tool aims to help therapists by facilitating the design and control of the training scenario. This ambitious goal requires multiple tests and validations (see Section~\ref{sez:conclusion}).
In this work, we focus on validating the personalization approach, demonstrating that:
\begin{enumerate}
    \item The proposed scenario personalization is able to define distinct sets of scenarios according to different user profiles;
    \item Given a $CD_{score}$ and the feature constraints of a certain user profile, the proposed scenario personalization is able to define scenarios that differ significantly, enabling ecological training environments.
\end{enumerate}

To this aim, we have defined four different synthetic user profiles describing their difficulties in the general task of crossing a street, which will represent the feature constraints. These profiles have the mere scope to allow us to simulate some of the characteristics that may occur in the ASD spectrum, but by any means must be considered neither exhaustive of the broad spectrum of individual differences, resulting in extremely complex profiles, nor as accurate of the overall spectrum of difficulties that may be registered in the described situation. 

The weights utilized in this study are determined based on our expertise with respect to the target user perception, and the actual values are empirically derived. These weights are detailed in Table~\ref{tab:weights}.
Point 1) of the evaluation also aims to verify qualitatively that the weighted scenarios are meaningful for the specific user profile. The following are the syntactic user profile descriptions.

\begin{itemize}
\item \textbf{Profile 1: Sound hypersensitive} - Unable to filter out loud noises, may freeze at the sound of the ambulance siren, may decide not to cross even when it is safe, as the sensory overload may obstruct the person's ability to assess and execute the necessary procedures to complete the action.\\
\textit{Relevant challenges}: Presence of loud noises.

\item \textbf{Profile 2: Excessively focused attention on a detail} - Focuses on the duration of the timer, unable to shift attention from that detail, does not cross, even though the situation is safe, because the timer dynamic locks the person's attentional resources and makes it difficult to be able to assess if the amount of time shown is enough to cross safely.\\
\textit{Relevant challenges}: Presence of the traffic light with a functioning timer, different lengths of crossings, and volumes of noise must be incremental.

\item \textbf{Profile 3: Social anxiety} - Avoids interaction with other pedestrians, crosses only during less crowded moments, possibly even if less safe.\\
\textit{Relevant challenges}: Presence of people, medium or long crossing, lack of safety.

\item \textbf{Profile 4: Intermittent attention} - Registers difficulties in maintaining the focus of attention for an extended amount of time, consequently registers difficulties in processing the complexity of the presented situation, in organizing the stimuli, and deciding on whether it is safe to cross or not.\\
\textit{Relevant challenges}: Longer crossing, medium presence of vehicles.

\end{itemize}

\begin{table}[t]
\centering
\caption{Mapping of personalization weights to scenario parameters and cognitive skill}
\label{tab:weights}

\begin{tabular}{c|ccccccc}
          & $\omega_1$ & $\omega_2$ & $\omega_3$ & $\omega_4$ & $\omega_5$ & $\omega_6$ & $\omega_7$ \\ \hline
\rowcolor[HTML]{EFEFEF} 
Profile 1 & 2          & 2          & 2          & 2          & 5          & 5          & 3          \\
Profile 2 & 3          & 1          & 3          & 3          & 3          & 3          & 3          \\
\rowcolor[HTML]{EFEFEF} 
Profile 3 & 2          & 2          & 5          & 5          & 2          & 2          & 2          \\
Profile 4 & 5          & 2          & 2          & 3          & 2          & 2          & 2         
\end{tabular}
\end{table}

\subsection{Measuring feature variation in scenarios with consistent difficulty}
We aim to examine how the features vary between scenarios of consistent difficulty. In particular, we are interested in determining whether each feature within a set of scenarios exhibits an even distribution of its possible values. For example, if a feature $\phi_i$ can take values in $\{0,0.5,1\}$ and we consider a set of scenarios, we define it as varying  ``as much as possible" if each value appears exactly $\frac{1}{3}$ of the time. 
In other words, based on this approach, for each feature $\phi_i$ that can assume $p_1, \dots, p_n$ values there is an ideal distribution $\overline{\phi_i}$ for its values, which is defined as:
\begin{equation}
    \mathbb{P}(\overline{\phi_i} = p_k) = 1/n \qquad \forall i =1,\dots,n.
\end{equation}
Thus, we use the divergence between the probability distributions to estimate the variation of features and ensure the generation of distinct scenarios. To do so, we exploit the \emph{Jensen–Shannon divergence}~\cite{lin1991divergence} ($JSD$ for short) between the distribution of a set of scenarios with consistent difficulty and its ideal distribution. 

To ease the visualization of this analysis, to estimate the variance $\mathcal{V}$ of a feature $\phi_i$ we use the following formula: 
\begin{equation}
    \mathcal{V}(\phi_i) = 1 - JDS(\phi_i,\overline{\phi_i}).
\end{equation}
Since $JDS$ varies between $0$ and $1$, $\mathcal{V}$ value is higher if the divergence is low (or rather, the distribution of $\phi_i$ values is good), and vice-versa. 

By computing $\mathcal{V}$ for each feature across all score values, we visualize the results using a multi-line graph (one per feature), providing a quantitative and intuitive representation of the features variation across different scores. In an ideal scenario, each lines would remain constant at value of $1$, indicating perfect uniformity in value distribution.

\section{Results}\label{sez:results}

For Profile 1, the challenge of the presence of loud noise has been translated in selecting scenarios including at least one feature ssd, i.e. church bell, helicopter or car waiting red lights. Fig.\ref{fig:pr1} illustrates the number of relevant scenarios clustered for the difficulty level. We can observe that once the constrain is applied, the number of scenarios does not decrease significantly for each $CD_{score}$, indeed 87,5\% of the scenarios includes the challenges considered relevant for profile 1.
Considering how the features vary between scenarios of consistent difficulty, we can observe that $\mathcal{V}(\phi_i)$ $\forall i$ rapidly approach a plateau near 1, indicating scenario diversity. Notice that among the axis key constrained features are not reported since they are fixed, and considering their variance is not meaningful for this evaluation.

For profile 2, feature constrain are represent by the presence of the traffic light with a functioning timer and the different lengths of crossings. The fact that volumes of noise must be incremental among the difficulty level has been translated into setting volumes under a certain threshold (minor of 33 \%) for easy level, among a certain range for middle range (grather that 33\% and minor than 66\%), and over a certain threshold for high level (greater that 66\%), then the results are depicted in Fig.\ref{fig:pr2} with three different charts, for easy/medium/high level of the training.
Despite this heavy restriction (for instance, only 2,7\% of the scenarios have the requirements for the easy level), we can positively observe that an important number of scenarios is present in each $CD_{score}$. This is also true for the middle and the high level. 
Also in this case, we can observe that the free feature tends to 1, ensuring a wide variety of scenarios within the same $CD_{score}$.

Profile 3 is characterized by many constraints. Indeed, scenario features require setting many people, medium or long crossing, and the lack of safety has been translated into a high number of vehicles and broken traffic light. To respect all the constraints, we can observe in Fig.\ref{fig:pr4}(top) that the minimum selectable $CD_{score}$ is 0.3. This evaluation is essential in our future work to set properly easy/middle/high level according to the $CD_{score}$ tailored for each person.

Finally, scenarios of profile 4 must present longer crossing and a medium presence of vehicles. Also in this case, we can observe in Fig.\ref{fig:pr4}(bottom) that after $CD_{score} = 0.5$ all the consistent difficulty score include a significative number of scenarios with a peak in the medium consistent difficulty score. The scenarios variance again shows a balanced use of all features value, converging towards 1 after $CD_{score}=1$.

Generally, across all profiles, we can observe that below a certain threshold (around 0,2) the variation of the feature is present ($\mathcal{V}(\phi_i)=0$ $\forall i$). The value of the threshold depends on the user profile, i.e., on the weights used to define the scenarios. The same observation is valid for the tails of the scenarios' variance plots, indicating that over a certain $CD_{score}$ (the specific threshold depends on the users' profile), the variation of the features is not ensured and then the proposed scenarios are quite similar among them.

\section{Conclusions and future works}
\label{sez:conclusion}

In this work, we proposed a method for personalizing urban training scenarios specifically tailored for neurodivergent individuals, allowing them to experience scenarios of increasing complexity. The proposed approach was evaluated using four synthetic user profiles, accompanied by a metric designed to assess the variance of a single feature within scenarios of equivalent difficulty levels.

The results demonstrate that our method effectively generates a substantial number of diverse scenarios across multiple difficulty levels. Importantly, significant variance is maintained within the same difficulty category, particularly for features that were not explicitly constrained by the specific user profile. This indicates the method's capability to deliver a rich spectrum of scenarios, offering valuable variability while maintaining targeted constraints.

Although preliminary, this research represents a critical step towards establishing precise thresholds for easy, medium, and difficult scenario classifications based on derived difficulty scores. Future work will involve collaboration with neurodivergent participants to accurately profile their responses and perceptions of difficulty. One possible approach is to propose a limited series of scenarios with significant differences to both the therapists and the users, collecting feedback on calibration. Subsequently, similar scenarios $CD_{score}$ will be presented to validate the consistency of the challenge at a fixed difficulty level. This feedback will also help to refine the assessment criteria, in order to achieve the ambitious goal of developing a semi-supervised personalization mechanism.


Moreover, different evaluation metrics beyond variance will be explored, searching for metrics that highlight scenarios diversity under different perspectives. Finally, computational complexity is not expected to pose a major challenge in this study, but are conducting stress tests using ``large" feature spaces to better understand the scalability and limitations.


\bibliographystyle{IEEEtran}
\bibliography{biblio}
\end{document}